\documentclass[aip,jcp,amsmath,amssymb,floatfix,reprint,citeautoscript,articletitle=false,noeprint]{revtex4-1}
\usepackage[english]{babel}
\usepackage{color}
\usepackage{graphicx}
\usepackage{wrapfig}
\usepackage[caption=false]{subfig} %
\usepackage{amsmath,amssymb,bm}
\usepackage[version=3]{mhchem}
\usepackage{verbatim}
\usepackage{multirow}
\usepackage{dcolumn}
\usepackage{float}
\usepackage{nicefrac}
\usepackage{siunitx}
\DeclareSIUnit[number-unit-product = {\,}]{\cal}{cal}
\DeclareSIUnit[number-unit-product = {\,}]{\kcal}{\kilo\cal}
\DeclareSIUnit[number-unit-product = {\,}]{\atomicunit}{a.u.}
\usepackage[colorlinks,allcolors=black,citecolor=blue,urlcolor=blue]{hyperref}

\newcommand{\zundel}{\cf{H5O2+}}

\begin{document}
\title{%
Converged
Colored Noise Path Integral Molecular Dynamics Study of the Zundel Cation 
down to Ultra-low Temperatures
at Coupled Cluster Accuracy}
\author{Christoph Schran}
\thanks{These authors contributed equally.}
\affiliation{Lehrstuhl f\"ur Theoretische Chemie,
  Ruhr-Universit\"at Bochum, 44780 Bochum, Germany}
\author{Fabien Brieuc}
\thanks{These authors contributed equally.}
\affiliation{Lehrstuhl f\"ur Theoretische Chemie,
  Ruhr-Universit\"at Bochum, 44780 Bochum, Germany}
\author{Dominik Marx}
\affiliation{Lehrstuhl f\"ur Theoretische Chemie,
  Ruhr-Universit\"at Bochum, 44780 Bochum, Germany}
\date{\today}

\begin{abstract}
\setlength\intextsep{0pt}
\begin{wrapfigure}{r}{0.5\textwidth}
  \hspace{-1.5cm}
  \includegraphics[width=0.5\textwidth]{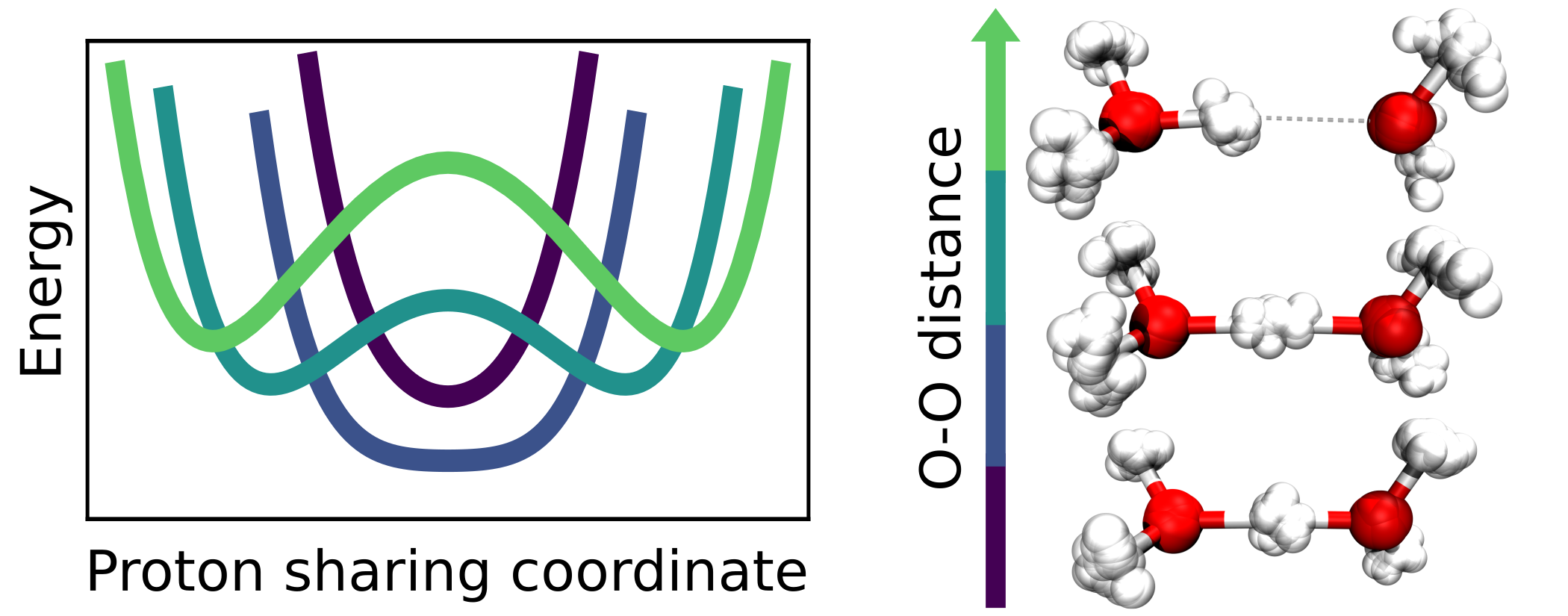}
\end{wrapfigure}
        For a long time, performing
        converged
        path integral simulations at ultra-low,
        but finite temperatures of 
        a few Kelvin has been 
        a nearly impossible task.
        However, recent developments in advanced colored noise thermostatting 
        schemes for path integral simulations, namely the Path Integral
        Generalized Langevin Equation Thermostat (PIGLET) and the Path 
        Integral Quantum Thermal Bath (PIQTB), have been able to greatly
        reduce the computational cost of these simulations, thus making 
        the ultra-low temperature regime accessible
        in practice. 
        In this work, we investigate the influence of these 
        two thermostatting schemes 
        on the description of 
        hydrogen-bonded 
        systems at temperatures
        down to a few Kelvin as encountered, for example, in helium 
        nanodroplet 
	     isolation or tagging photodissociation spectroscopy experiments. 
        For this purpose, we 
        analyze
        the prototypical hydrogen bond in
        the Zundel cation (\zundel{}) as a function of 
        both,
        oxygen-oxygen 
        distance and temperature in order to elucidate how the 
        anisotropic quantum delocalization and, thus, the 
        shape of the shared proton adapts
        depending on the donor-acceptor distance. 
        The underlying electronic structure of the Zundel cation is described 
        in terms of Behler's Neural Network Potentials of 
        essentially converged Coupled Cluster accuracy, 
        CCSD(T*)-F12a/AVTZ. 
        In addition, the performances of the PIQTB and PIGLET methods 
        for energetic, structural, and 
        quantum 
        delocalization properties are
        assessed and 
        directly compared.
        Overall, our results emphasize the validity 
        and practical usefulness of these two modern
        thermostatting 
        approaches for 
        path integral simulations 
        of hydrogen-bonded systems even at
        ultra-low temperatures.
\end{abstract}

\maketitle

\section{Introduction}
\label{sec:intro}
Experimental techniques operating at ultra-low temperatures 
such as helium nanodroplet isolation~\cite{Toennies1998/10.1146/annurev.physchem.49.1.1,
Toennies2004/10.1002/anie.200300611,Toennies2013/10.1080/00268976.2013.802039}
and tagging photodissociation~\cite{Okumura1986/10.1063/1.451079,
Roithova2016/10.1021/acs.accounts.5b00489}
spectroscopy
experiments have provided many insights on the nature 
and properties of fragile molecules and clusters that cannot be studied 
in the gas phase on their own.
Computational studies complement these experiments and help elucidating
some of the microscopical mechanisms at work
as, for example, shown for the dissociation of
microsolvated HCl/water clusters~\cite{Gutberlet2009/10.1126/science.1171753,Forbert2011}
or understanding of proton transfer in protonated
water clusters~\cite{Wolke2016/10.1126/science.aaf8425}.
However, at these ultra-low temperatures, nuclear quantum effects (NQEs), 
such as quantum delocalization 
due to zero-point motion 
and tunneling
need to be taken into account, if qualitative or even quantitative 
conclusions are of interest.
In such cases, simulation techniques based on the path integral (PI) formalism 
of quantum mechanics, such as PI molecular dynamics (PIMD)~\cite{
Callaway_1982_a, Parrinello_1984_a,
Tuckerman2010},
are a reliable tool to explicitly include NQEs.
These methods use the discretized imaginary time path integral formulation of 
the quantum partition function allowing, in principle, to obtain the exact 
quantum thermodynamic properties of the system~\cite{Feynman1965,Chandler1981,Ceperley1995}.
In practice, properties converge towards their exact quantum 
values when the discretization of the path integral $P$
(usually called number of replicas, beads or Trotter index)
is large enough.
However, the number of replicas to reach this convergence
increases unfavorably upon lowering the temperature,
thus making it especially complicated to
treat systems at temperatures of a few Kelvin.
A wealth of 
different
methods have been proposed to accelerate PIMD simulations 
that are based on 
distinct 
approaches such as 
higher-order Trotter discretization schemes~\cite{Takahashi1984,
Li1987,Suzuki1995,Chin1997,Perez2011,Kapil2016}, 
actions based on the pair density matrix~\cite{Pollock1984,Ceperley1995}, 
ring polymer contractions~\cite{Markland2008,Marsalek2016,John2016,Cheng2016}, 
or colored noise thermostatting~\cite{Ceriotti2011,Brieuc2016a} 
as well as post-processing or reweighting schemes~\cite{Jang2001,Yamamoto2005,Poltavsky2016}
to name but a few. 
Among these approaches, the use of colored noise thermostats is
able to significantly reduce the number of replicas required to reach the 
convergence of PIMD simulations~\cite{Ceriotti2011,Brieuc2016a} which 
is necessary if one wants to reach ultra-low temperatures~\cite{Uhl2016}.
In particular, it has been shown that the Path Integral Generalized Langevin 
Equation Thermostat (PIGLET)~\cite{Ceriotti2012}, can be used to accelerate
the convergence by around two orders of magnitude even in the complicated
case of the 
highly fluxional and thus 
utmost anharmonic 
protonated methane molecule (\cf{CH5+})
at ultra-low 
temperature~\cite{Uhl2016}. 
Another colored noise thermostatting scheme to accelerate PIMD
simulations, based on the same principle as PIGLET but using a 
different colored noise thermostat, called Path Integral Quantum 
Thermal Bath (PIQTB), has recently been developed~\cite{Brieuc2016a}.
However, the performance of this method at ultra-low temperatures
as well as its comparison to PIGLET simulations 
remain unknown up to now.

In the present work we 
systematically scrutinize these two colored noise thermostatting schemes 
by studying 
NQEs on
the prototypical hydrogen bond in the Zundel cation~\cite{Tuckerman1997,
Spura2015}
as a function of temperature
down to the order of one~Kelvin. 
Moreover, 
varying the donor-acceptor distance with the help of suitable restraints
allows us to continuously transform a strong and centered
hydrogen bond 
to a weak 
and very asymmetric 
one. 
This enables us to quantify if and to which extent PIGLET and PIQTB 
artificially change the shape of nuclei that are subject to 
anisotropic quantum delocalization, such as protons in hydrogen bonds
of different length. 
For this purpose, we 
analyze 
how the PIGLET and PIQTB methods perform for various energetic, structural
and quantum delocalization properties, thus for the first time
comparing these two methods directly with respect
to standard PIMD simulations
that have been converged by using very large replica numbers $P$. 

The outline of the paper is as follows: 
We first describe the relevant aspects of the methodology 
with a focus on the similarities and differences of PIGLET and PIQTB
in Sec.~\ref{sec:methods} and provide the computational details
in Sec.~\ref{sec:comp-det}.
In the first two parts of Sec.~\ref{sec:res}, we compare
the path integral convergence of energetic and structural properties 
of the Zundel cation.
Afterwards, delocalization properties are addressed
in the last two parts of Sec.~\ref{sec:res} where we investigate,
in particular, the convergence of the quantum delocalization
and, in order to draw some general conclusions, the anisotropy
of this delocalization as a function of the 
heavy atom distance.
Finally, we conclude and 
suggest
some directions for future work.

\section{Methods: Accelerated Path Integral Molecular Dynamics}
\label{sec:methods}
Although stochastic thermostatting has a long tradition in 
PIMD to render sampling of the path integral ergodic, 
stochastic thermostats using colored noise have been explored only fairly recently.
Initially, these colored noise methods have been developed to 
efficiently 
emulate
NQEs in standard MD simulations, 
thus avoiding the computationally much more demanding PIMD simulations altogether. 
Two different 
underlying principles
have been 
exploited to this end, namely 
the Generalized Langevin Equation (GLE)~\cite{Ceriotti2009b} and the Quantum Thermal Bath (QTB)~\cite{Dammak2009}. 
The main idea behind both of these approaches is to achieve a frequency-dependent
thermalization of the system so that the average energy of a vibrational mode of 
angular frequency $\omega$,
\begin{align}
\langle E(\omega)\rangle = \frac{\hbar\omega}{2}\coth\left(\frac{\beta\hbar\omega}{2}\right)
,
\label{eq:Ew}
\end{align}
is given by the exact quantum expression in the harmonic 
approximation. 
In the GLE case, this is achieved using the generalized Langevin equation 
rewritten in a Markovian form by introducing a set of additional degrees of 
freedom (DoFs) that represents the quantum bath~\cite{Ceriotti2010b,Ceriotti2015}.
Although the equations of motion are then Markovian, the coupling between the 
physical and the additional DoFs leads to a non-Markovian process for 
the physical variables~\cite{Ceriotti2010b,Ceriotti2015}.
The introduction of these additional DoFs allows for great flexibility 
so that the GLE can actually be used for different purposes such as 
optimal sampling~\cite{Ceriotti2009a}, selective normal 
mode excitation~\cite{Ceriotti2010a,Ganeshan2013} 
and, of course, frequency-dependent thermostatting~\cite{Ceriotti2009b,Ganeshan2013} 
in particular to include NQEs.
However, this flexibility comes at the price of a large number of parameters 
which need to be carefully optimized in order to accurately reproduce the 
energy distribution of Eq.~(\ref{eq:Ew}).
This is done, in practice, using a highly complex and non-trivial fitting 
procedure
to parameterize the GLE approach~\cite{Ceriotti2010b}. 
In the QTB case, the target energy distribution is enforced using a modified Langevin 
thermostat where the Gaussian white noise is replaced by a colored noise~\cite{Dammak2009}.
The
standard
Langevin thermostat introduces a friction force, with a friction coefficient
$\gamma$, and a random force $R$ so that the equation of motion for a DoF
$i$ is 
given by 
the Langevin equation, 
\begin{align}
        \dot{\text{p}}_i = -\nabla_i V(\{\text{q}\}) - \gamma\text{p}_i + \sqrt{2m_i\gamma}R_i.
\end{align}
In the case of usual classical nuclear dynamics at constant temperature $T$, 
the random force is a Gaussian white noise whose power
spectral density $I_{R_i}$ is given by the classical fluctuation-dissipation 
theorem~\cite{Kubo1966} 
and is equal to the thermal energy 
$k_{\rm B}T$ so that the 
autocorrelation function of $R_i$ 
satisfies 
\begin{align}
\langle R_i(t)R_i(t+\tau) \rangle = k_{\rm B}T\delta(\tau).
\end{align}
In the QTB case, the Gaussian white noise is replaced by a colored noise which thus
has a frequency-dependent power spectral density $I_{R_i}(\omega)$ that is equal to 
the target energy distribution $\langle E(\omega) \rangle$ of Eq.~(\ref{eq:Ew}) 
according to the quantum fluctuation-dissipation theorem as formulated by Callen
and Welton~\cite{Callen1951}.
The autocorrelation function of $R_i$, 
\begin{align}
\langle R_i(t)R_i(t+\tau) \rangle = \int_{-\infty}^{\infty} I_{R_i}\,
\text{e}^{-i\omega\tau}\,\text{d}\omega
, 
\end{align}
is then obtained using the Wiener-Khinchin theorem.
One very appealing aspect of the QTB is its relative simplicity compared to the 
GLE thermostat.
Indeed, the method only depends on a few parameters and relies on a simple Langevin
thermostat that is easy to implement and is already available in most MD codes.
In practice, one only needs to generate random forces having the 
correct
power spectrum which can easily be done on-the-fly 
using a moving average (or finite impulse response, FIR) filter 
as proposed earlier~\cite{Barrat2011}.
Both the QTB and GLE methods are able to include part of the quantum 
effects, such as zero-point energy, in a computationally inexpensive way
and have been successfully applied to various systems~\cite{Ceriotti2010d,
Hassanali2012,Dammak2012,Calvo2012a,Calvo2012b,Bronstein2014,Bronstein2016,
Bronstein2017,Schaack2018}. 
However, the use of these thermostats is restricted to weakly anharmonic systems.
Indeed, the energy distribution of Eq.~(\ref{eq:Ew}) is formally valid in 
the harmonic case only, but the main limitation of these methods in anharmonic 
cases is that they are prone to the problem of zero-point energy leakage (ZPEL).
Due to anharmonic couplings between the normal modes, part of the energy is
transferred from the high frequency modes to the low frequency ones, leading a the 
wrong energy distribution~\cite{Ceriotti2010b,Bedoya2014,Hernandez2015,Brieuc2016b}.
Even though it is possible to limit the effect of ZPEL by enforcing a strong
coupling between the system and the thermostat~\cite{Ceriotti2010b,Brieuc2016b},
it is impossible to know how accurate the obtained results are when studying
unknown systems.

Beyond approximately including NQEs in a classical MD framework via thermostatting, 
these two colored noise schemes have been shown to be particularly useful 
in combination with PIMD, where ergodic thermostatting is 
fundamentally required~\cite{Tuckerman2010}. 
Within the PIMD framework, GLE- and QTB-based methods 
can be 
systematically 
converged towards the
exact results by increasing the number of beads~\cite{Ceriotti2011,Brieuc2016a}. 
Parts of the NQEs are already included via the thermostat,
which implies immediately that
less replicas compared to canonical PIMD simulations are required to reach convergence~\cite{Ceriotti2011,Brieuc2016a}.
Thus, the use of colored noise thermostats is able to greatly accelerate the 
convergence of standard PIMD~\cite{Ceriotti2011,Brieuc2016a}.

In the framework of accelerating Trotter convergence of PIMD simulations, the
main idea 
of colored noise thermostats
remains to achieve a frequency-dependent thermalization of the 
system to impose the right quantum energy distribution in the harmonic approximation.
However, this energy distribution now applies to all the normal modes of the 
ring polymer and depends on the number of replicas $P$ used in the PIMD simulation.
Let us consider the simple case of a one dimensional harmonic oscillator of 
angular frequency $\omega$. 
Its average energy can be expressed as
\begin{align}
\langle E(\omega)\rangle_P = \frac{1}{P} \sum_{k=1}^P
\frac{\omega^2}{\omega_k^2} \langle \tilde{E}(\omega_k)\rangle_P 
, 
\label{eq:Ewk}
\end{align}
where $\langle \tilde{E}(\omega_k)\rangle_P$ is the average energy of the 
ring polymer normal mode $k$ of angular frequency $\omega_k$;
here $\omega_k^2 = \omega^2 + 4\,\omega_P^2\sin^2\left((k-1)\pi/P\right)$ with $k=1, \dots P$.
In the standard PIMD case, the ring polymer is classically thermalized at the 
temperature $T\times P$, so that $\langle \tilde{E}(\omega_k)\rangle_P = k_\text{B}T \times P$.
When the number of replicas $P$ increases, the average energy of the harmonic
oscillator converges towards its exact quantum value given by Eq.~(\ref{eq:Ew}).
Colored noise thermostats can be used in order to ensure that 
$\langle E(\omega)\rangle_P$ is given by the exact quantum expression 
for any value of $P$. 
This is achieved by enforcing an energy distribution 
$\langle \tilde{E}(\omega_k)\rangle_P$ 
among the ring polymer normal modes that is 
the
solution of the following equation
\begin{align}
\frac{1}{P} \sum_{k=1}^P \frac{\omega^2}{\omega_k^2} 
\langle \tilde{E}(\omega_k)\rangle_P = 
\frac{\hbar\omega}{2}\coth\left(\frac{\beta\hbar\omega}{2}\right)
, 
\end{align}
which is directly obtained by combining Eqs.~(\ref{eq:Ew}) and (\ref{eq:Ewk}).
This equation can be solved numerically for any values of $P$ and the obtained 
energy distribution $\langle \tilde{E}(\omega_k)\rangle_P$ can then be 
enforced using either the GLE thermostat, leading to the PIGLET
method~\cite{Ceriotti2011,Ceriotti2012}, or the QTB, leading to 
the PIQTB method~\cite{Brieuc2016a}.
In the latter case, each ring polymer normal mode is submitted to 
a random force that is a colored noise whose power spectral density is
the target energy distribution $\langle \tilde{E}(\omega_k)\rangle_P$ in 
the same spirit as 
in plain
QTB.
Both methods are able to include part of the quantum fluctuations
(mostly the harmonic part) via the thermostat so that the number $P$ of
replicas required to Trotter-converge the PIMD simulation is significantly reduced.

For PIGLET, we refer to a recent publication~\cite{Uhl2016}
concerning its details and  implementation in the 
\texttt{CP2k} program package~\cite{cp2kDev,Hutter2014} 
as well as its performance assessment in the realm of ultra-low temperature simulations, 
whereas the PIQTB technique as such and its implementation in \texttt{CP2k}
are described in Sec.~I of the Supporting Information.

\section{Computational details}
\label{sec:comp-det}
All simulations have been carried out using the 
\texttt{CP2k} 
simulation package~\cite{cp2kDev,Hutter2014}.
In particular, the standard PIMD
reference simulations
(where Trotter convergence is achieved by using very large $P$ values) 
have been carried out using the 
standard path integral Langevin equation (PILE)
thermostat~\cite{Ceriotti2010c} with a friction parameter for the centroid 
mode of $\gamma =$ \SI{1}{\pico\second^{-1}} 
whereas the 
friction coefficients for 
the other modes 
are given by
$\gamma_{k} = \tilde{\omega}_k$ with $k=2,...,P$ 
where 
$\tilde{\omega}_k$ 
are 
the usual angular frequencies of the free
particle
ring polymer normal modes 
given by $\tilde{\omega}_k=2\omega_P\sin\left((k-1)\pi/P\right)$.
In terms of \texttt{CP2k} parameters this setup corresponds
to $\tau=$\SI{1000}{\femto\second} and $\lambda = 0.5$.
The PIGLET simulations have been performed using the matrices (containing all 
the GLE parameters) as developed in Ref.~\citenum{Uhl2016} to study \cf{CH5+} at ultra-low
temperatures.
For PIGLET, we are thus limited to a maximum of $P=$~32, 64 and 512 
beads at $T=$~100, 20 and 1.67~K, respectively.
The PIQTB method 
has been implemented by us
into the \texttt{CP2k} simulation package.
In the same spirit as for the PILE thermostat, every ring polymer normal mode 
is coupled to its own QTB thermostat except for the centroid ($k=1$) mode which is 
attached to a classical Langevin thermostat.
The value of the friction coefficient of the centroid normal mode 
is set to $\gamma =$~\SI{20}{\pico\second^{-1}} and the friction coefficients for the other modes 
($k=2,...,P$) are scaled so that $\gamma_k^2=\gamma^2 + \lambda^2\,\tilde{\omega}_k^2$ 
with $\lambda=0.2$.
The angular cutoff frequency we use for the centroid mode is
$\omega_{\text{cut},1}=$\SI{1100}{\radian.\pico\second^{-1}} and the cutoff frequencies
for the other modes are scaled so that 
$\omega_{\text{cut},k}^2=\omega_{\text{cut},1}^2+\lambda_\text{cut}^2\,\tilde{\omega}_k^2$
with $\lambda_\text{cut}=1.5$.
In terms of \texttt{CP2k} parameters this setup corresponds 
to $\tau=$\SI{50}{\femto\second}, $\lambda = 0.2$,
$\tau_\text{cut}=$\SI{0.9}{\femto\second} and $\lambda_\text{cut} = 1.5$.
In the following, all kinetic energies using PIQTB thermostatting are obtained using 
the modified virial estimator introduced in Ref.~\citenum{Brieuc2016a}.
More details about the implementation of PIQTB in \texttt{CP2k}, the choice of
the parameters and the kinetic energy estimator are given in the
Supporting Information.
All reported PI
simulations were propagated for \SI{0.5}{\nano\second}
using a
formal molecular dynamics
timestep of \SI{0.25}{\femto\second}
where \SI{10}{\pico\second} at the beginning of each
simulation were discarded as equilibration from the
starting configuration.

The electronic structure of the Zundel cation is in principle accessible
up to Coupled Cluster theory by on-the-fly electronic structure
calculations as impressively shown in the pioneering study of
Ref.~\citenum{Spura2015} made possible due to modern
Car-Parrinello-like approaches~\cite{Kuehne2007}.
However, full convergence of the electronic structure and
path integral discretization, especially at ultra-low temperatures,
are still out of scope for these approaches.
Energies and forces of the Zundel cation were,
therefore,
obtained
using a Neural Network Potential~\cite{Behler2007/10.1103/PhysRevLett.98.146401,
Behler2017/10.1002/anie.201703114} fitted to 
CCSD(T*)-F12a/AVTZ
reference data that has been developed similarly as described in
Ref.~\citenum{Schran2018/10.1063/1.4996819}.
As demonstrated in the Supplemental Material therein~\cite{Schran2018/10.1063/1.4996819},
this machine learning approach provides electronic-structure-based interactions
and thus the global potential energy surface (PES) of the protonated water dimer
at the essentially converged Coupled Cluster level.
This allows us similarly as recently demonstrated for a different machine learning
technique~\cite{Tkatchenko2018} to run electronically converged simulations at
the gold standard of modern quantum chemistry.

In order to systematically investigate the response of the 
shape of the proton in the hydrogen bond with respect
to the intermolecular distance, the
symmetric Zundel complex \mbox{[H$_2$O$\cdots$H$\cdots$OH$_2$]$^+$}  
was restrained by 
applying two identical external 
three-dimensional isotropic
harmonic potentials that exclusively act on
every bead of
the 
two oxygen atoms
in the laboratory fixed coordinate system. 
After careful testing, the spring constants of the two harmonic 
potentials were set to \SI{0.5}{\hartree\bohr^{-2}} to
enforce the desired heavy atom separation while still allowing
for sufficient fluctuations.
If not specified differently, the
shared proton was initalized between the oxygen atoms
using the OH~distance of the equilibrium structure, 
thus closer to one of the oxygen atoms for the larger O-O
separations.
Note that the respective observables in the next section are reported
as a function of the resulting quantum expectation value of the donor-acceptor distance,
being the average oxygen-oxygen distance $\langle r_\text{OO} \rangle $ 
as obtained from the PI simulations, at the given temperature.

\section{Results and Discussion}
\label{sec:res}
In the following we carry out a detailed comparison of the performance of the PIQTB 
and PIGLET methods to study the Zundel cation down to ultra-low temperatures.
In the first two parts, we focus on the convergence behavior of the two methods 
as compared to standard PIMD for several quantities of interest
such as average energies and selected structural properties.
Then, in the last two parts, we study 
systematically
the impact of these thermostats 
on the delocalization properties of the atoms and, in particular, of the shared proton
as a function of the hydrogen bond length. 
\subsection{Energetic Properties}
\label{sec:ener}
\begin{figure*}[ht!]
\includegraphics[width=1.0\textwidth]{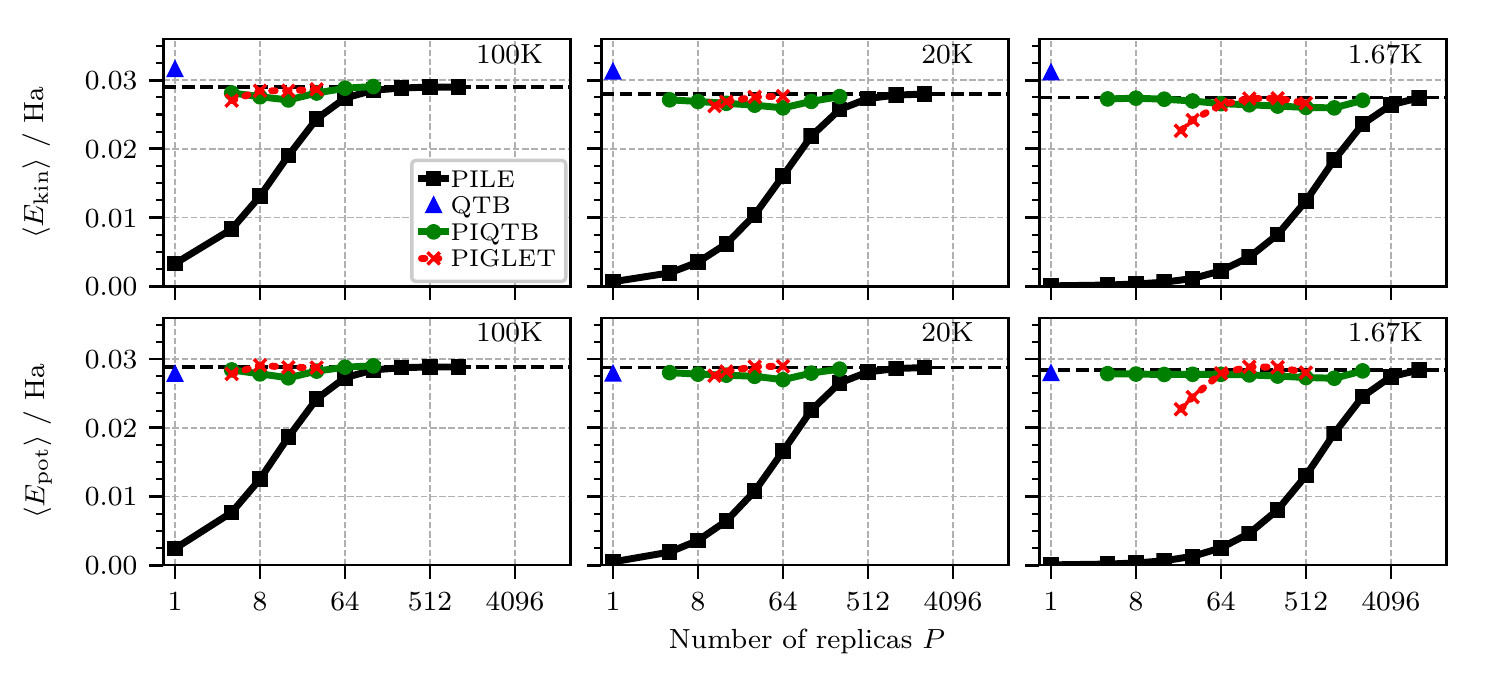}
\caption{Convergence of the average kinetic (top) and potential (bottom)
         energies 
of the Zundel cation, \zundel,
         with respect to the number of replicas $P$ at 
         $T=$~\SI{100}{}, \SI{20}{} and \SI{1.67}{\kelvin} as obtained 
         using the PILE (black squares), PIGLET (red crosses) and
         PIQTB (green circles) thermostats. 
         The blue triangles show
         the energies obtained with the bare QTB ($P=1$) thermostat, 
which corresponds to a classical MD simulation with QTB colored noise thermostatting. 
         The potential energies presented here are 
given 
         relative to the potential energy of the 
         equilibrium structutre of \zundel{}.
         Note that the statistical error bars as obtained from
         block averaging are smaller than the
         symbol size in all shown cases.
         }
\label{fig:conv-ener}
\end{figure*}

As mentioned above, colored noise thermostats have been designed to speed up the
convergence of PIMD simulations targeting specifically the convergence of the average
kinetic and potential energies.
We thus start here by comparing the convergence of these average energies
with respect to the number of path integral replicas $P$
as obtained from PILE, PIGLET and PIQTB simulations
of the Zundel cation at 
100, 20 and finally 
\SI{1.67}{\kelvin} which is depicted in Fig.~\ref{fig:conv-ener}.
Starting from classical MD simulations 
(obtained from PILE simulations in the $P=1$ limit), 
increasing the
number of path integral replicas in the canonical PILE simulations
leads to a monotonic increase of the average energies that finally 
converges to a constant 
plateau
value equal to the quantum average energy.

As expected, the number $P$ of beads required to reach this
convergence limit increases 
drastically 
when the temperature is decreased;
note the logarithmic $P$ scale in Fig.~\ref{fig:conv-ener}. 
In case of the PILE thermostat, around $128$ replicas are necessary 
at \SI{100}{\kelvin}, while at \SI{20}{\kelvin}, 
a Trotter number of approximately $P=512$ is needed.
Finally, at \SI{1.67}{\kelvin}, one needs to use at least $4096$
replicas in order to reach convergence.
This is a clear illustration of the limitation of standard PIMD to study 
systems at such ultra-low temperatures.
Additionally, one can see that the converged energies are
basically constant with temperature indicating that the system
is almost in its ground state already at \SI{100}{\kelvin}.
Thus the main contribution to the energies comes from zero-point energy 
whereas 
the thermal fluctuations are negligible compared to the quantum ones 
at all temperatures that we address in the present investigation. 

The use of colored noise thermostats is clearly accelerating the convergence of the 
energies as seen in Fig.~\ref{fig:conv-ener}.
Indeed, the average energies obtained by PIQTB and PIGLET are very close to their 
converged values even for rather low values of $P$. 
This is particularly striking for the PIQTB method at \SI{1.67}{\kelvin} 
where the energies are already equal to the converged PILE results 
for $P$ as low as $4$.
However, as it has already been shown for the PIGLET thermostat~\cite{Uhl2016},
the convergence of these methods can exhibit a non-monotonic behavior.
Here, this is especially noticeable in the PIQTB case 
at all three temperatures
where a small kink
is present in the convergence of the energies 
that shifts to higher values of $P$ when the temperature is lowered.

The PIGLET method exhibits this behavior only at \SI{1.67}{\kelvin},
where we obtain average energies for $P=512$ that are lower than the ones obtained for 
$P=256$.
Note that due to the complicated construction of the PIGLET matrices
we are limited to 512 replicas at \SI{1.67}{\kelvin} for PIGLET,
while the number of replicas for PIQTB can continuously be changed without 
difficulty.
The origin of this non-monotonic convergence is not totally clear but, as already
discussed in Ref.~\citenum{Uhl2016}, is 
most
probably related to ZPEL.
In any case, the overall convergence behavior of both PIQTB and PIGLET is rather
similar and we can see that both methods are very close to 
the converged value for almost
every number of beads, showing that these thermostats are significantly
improving the convergence 
of both, the kinetic and potential energy
even down to ultra-low temperatures.
Let us also note that even the bare QTB method (\textit{i.e.} PIQTB for 
$P=1$) is already giving 
quite reliable energy
estimates even though it
tends to systematically overestimate the kinetic energy.
Since both methods have been specifically designed in order to accelerate the 
convergence of the average energies, the next step is
to investigate how they perform for other physically relevant quantities
such as structural observables.

\subsection{Structural Properties}
\label{sec:struct}
Next, we
focus on the 
molecular
structure of the Zundel cation and study in particular
the distribution of three selected 
observables:
the OH distance between the
dangling hydrogen atoms and their closest oxygen atom, $r_\text{OH}$, 
the hydrogen bond angle between the shared proton and the two oxygens, 
$\angle_\text{OHO}$, and the proton sharing 
(or proton transfer) 
coordinate 
$\delta = r_{\text{O}_1\text{H}}-r_{\text{O}_2\text{H}}$ with
$r_{\text{O}_1\text{H}}$ and $r_{\text{O}_2\text{H}}$ being the distance between the
shared proton and oxygen O$_1$ and O$_2$, respectively.
In order to both,
reliably and
quantitatively study the convergence of the different 
observables $x$, 
we need to compare the 
entire distribution functions 
$\rho_P(x)$ obtained for a certain value of $P$ 
to the converged one, 
$\rho(x) = \lim_{P\to\infty} \rho_P(x)$, 
obtained with the PILE thermostat 
using the largest 
number of beads.
We chose to perform this comparison by computing the distance between 
$\rho_P$ and $\rho$ using a simple metric 
\begin{align}
    d(\rho_P,\rho) = \frac{\int_{-\infty}^{+\infty} \left| \rho_P(x)-\rho(x)\right|
                     \text{d}x}
                     {\int_{-\infty}^{+\infty}\rho_P(x) \text{d}x + \int_{-\infty}^{+\infty}\rho(x) \text{d}x} \label{eq:dP}
\end{align}
based on the absolute difference of the two distributions
normalized to a maximum distance of 
unity. 
The interested reader can find all distributions for $r_\text{OH}$ 
distances, the $\angle_\text{OHO}$ angle and the $\delta$ coordinate
along with distributions of a few more quantities in the 
Supporting Information.
\begin{figure*}
\includegraphics[width=1.0\textwidth]{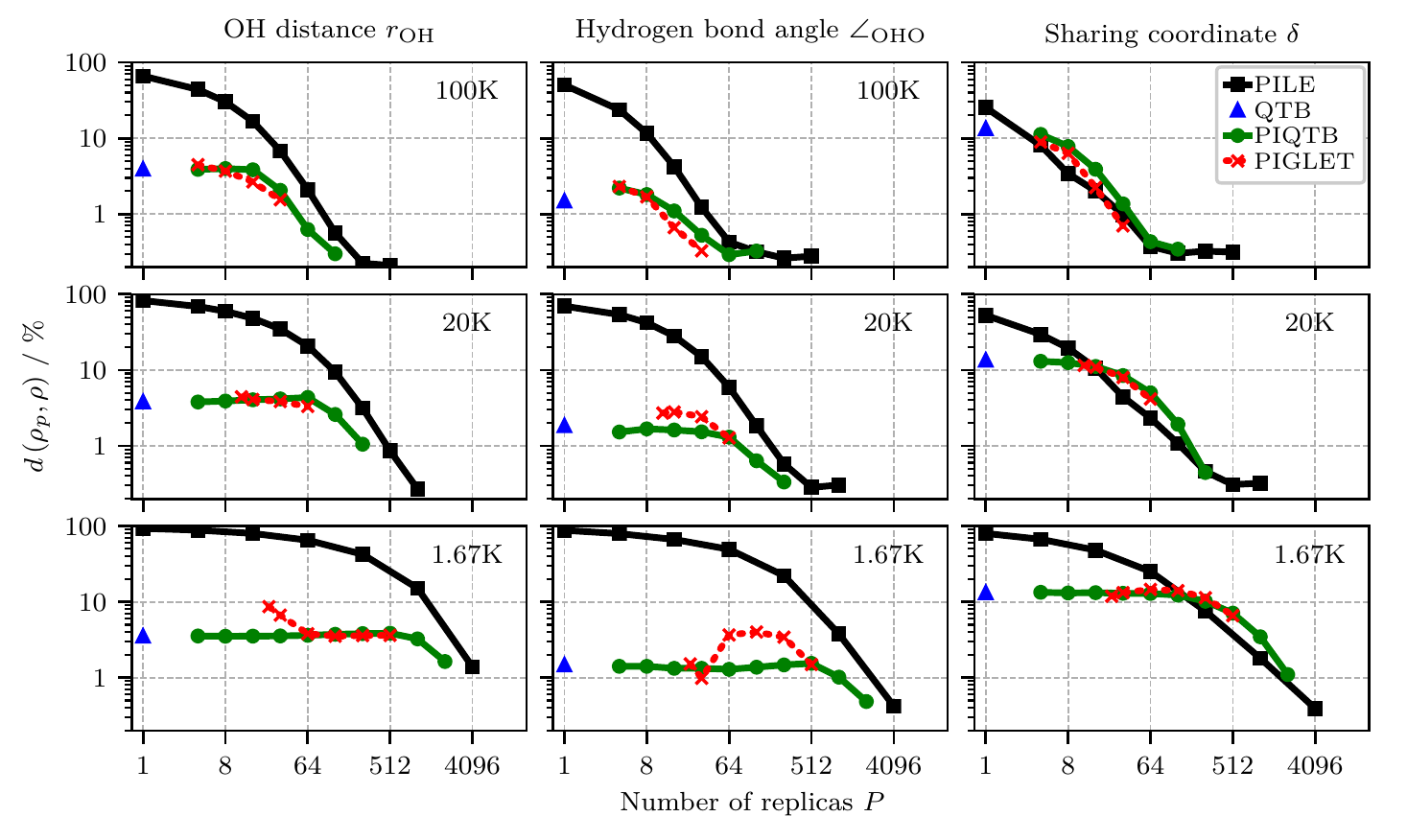}
\caption{Convergence 
of 
         the distribution functions of $r_\text{OH}$ distances (left column), 
         the $\angle_\text{OHO}$ angle (center column) and 
         of the proton sharing 
(transfer) 
coordinate $\delta$ (right column) 
of the Zundel cation, \zundel, 
with respect to the number of replicas $P$
         at $T=$~\SI{100}{}, \SI{20}{} and \SI{1.67}{\kelvin}
as obtained
         using the PILE (black squares), PIGLET (red crosses) and
         PIQTB (green circles) thermostats.
         The blue triangles show the results obtained with the bare QTB
         ($P=1$) thermostat.
         The measure of convergence as a function of $P$ is
defined in Eq.~(\ref{eq:dP}).
         Note that the statistical error bars are smaller than the
         symbol size in all shown cases.
    }
\label{fig:conv-struct}
\end{figure*}
The convergence of the three distributions
is compiled in Fig.~\ref{fig:conv-struct} 
by showing the distance $d(\rho_P,\rho)$ as a function of $P$.
In contrast to what has been found for the energies, 
the distributions of these structural observables 
converge 
more slowly to the 
$P\to \infty$ quantum distribution
even when using PIGLET or PIQTB, 
yet improved convergence is observed as before
considering
that a distance of 
$d(\rho_P,\rho) \approx 1-3$~\%
corresponds to statistically identical distributions.
Yet, much 
higher bead numbers are necessary for the PILE thermostat
to reach convergence 
in particular 
when the temperature is 
low. 
The convergence behavior of PIGLET and PIQTB 
follows a similar dependence
as PILE, 
but starts 
much
closer to the converged results
for  small numbers of replicas;
note the double-logarithmic scale in this figure.
As seen for the PILE thermostat, the number of beads necessary 
to reach convergence increases when the temperature decreases, 
but the convergence of PIQTB and PIGLET becomes
especially slow when reaching ultra-low temperatures.
For example, increasing $P$ up to $512$ has almost no effect on the convergence 
of the $r_\text{OH}$ distribution for PIQTB and PIGLET at \SI{1.67}{\kelvin}.
Nevertheless, for the distribution of OH distances and the hydrogen bond angle, 
the use of colored noise thermostats clearly accelerates the convergence 
as compared to PILE even down to ultra-low temperatures,
see Fig.~\ref{fig:conv-struct}.
Thus, a great gain in accuracy is already obtained compared to 
standard PIMD
when using a surprisingly small bead number in either PIGLET or PIQTB,
whereas ultimate convergence toward the $P\to\infty$ quantum limit 
sets in only at unpleasantly large Trotter numbers, in particular at low temperatures.

In case of the proton sharing coordinate $\delta$, however, both PIQTB and PIGLET
seem to have great difficulties to converge
to the exact result 
according to the right column in Fig.~\ref{fig:conv-struct}. 
This behavior is due to the fact that the potential associated with
the $\delta$ coordinate is highly anharmonic.
The top panel of Fig. \ref{fig:delta_doo} depicts 
cuts of the Coupled Cluster PES of \zundel\ 
showing
the potential energy profile
along the proton sharing coordinate for different fixed values of the oxygen distance
$r_\text{OO}$.
In our case, the average oxygen distance turns out to be
$\langle r_\text{OO} \rangle \approx \SI{2.415}{\AA}$ 
for all temperatures studied here and one can see that
the associated energy profile resembles a 
flat quartic potential at the corresponding value of $r_\text{OO} = 2.4$~\AA . 
Highly anharmonic potentials such as the 
pure
quartic well are one of the most 
difficult cases for accelerated colored noise thermostats
since the harmonic term is completely absent~\cite{Dammak2011a,Dammak2011b}.
It is thus not surprising that these thermostats show 
such a 
slow convergence of the distribution of $\delta$
in this particular case.
Note that due to the complex construction of the PIGLET matrices,
we are not able to reach full convergence for this method as discussed before
for the energetic properties.
Despite this general observation, we would like to point out
that in particular at the ultra-low temperature of 1.67~K, 
using either PIGLET or PIQTB leads to an improvement of one 
order of magnitude 
on the deviation $d(\rho_P,\rho)$ of the proton transfer coordinate 
w.r.t. its converged distribution 
over standard PILE simulations
in the small $P$ limit.

In conclusion, both the PIQTB and PIGLET methods
are able to accelerate the convergence
of most structural properties, although one needs to be careful when studying
highly anharmonic properties such as the proton sharing coordinate
which relates to an intermolecular distance.

\subsection{Nuclear Delocalization Properties}
\label{sec:delocalization}
Having confirmed that colored noise thermostats are able to accelerate 
the convergence of both, energies and structural properties even
down to ultra-low temperatures, we now investigate the effect of these thermostats
on the delocalization of the 
nuclei. 
The 
nuclear 
delocalization is directly related to the fluctuations of the position
operator 
\begin{align}
\sigma^2=\frac{1}{P}\sum_{s=1}^P\left< \left(\textbf{r}_s-
\langle\textbf{r}_s\rangle\right)^2\right>
\end{align}
which, in the path integral framework, can be decomposed into 
two additive contributions~\cite{Herrero2016}, 
\begin{align}
\sigma^2=\sigma_{\rm c}^2 + r_{\rm g}^2
, 
\end{align}
where the first term 
is the fluctuations of the centroid
of the ring polymer $\textbf{r}_{\rm c} = \sum_{s=1}^P \textbf{r}_s / P$ which
contains the thermal fluctuations.
The second term is the squared radius of gyration, 
\begin{align}
r_{\rm g}^2=\frac{1}{P}\sum_{s=1}^P\left<\left(\textbf{r}_s-
\textbf{r}_{\rm c}\right)^2 \right>
, 
\end{align}
that quantifies the spread of the ring polymer and, thus, is a direct measure 
of the quantum delocalization
at the given temperature.

\begin{figure}
\includegraphics[width=0.5\textwidth]{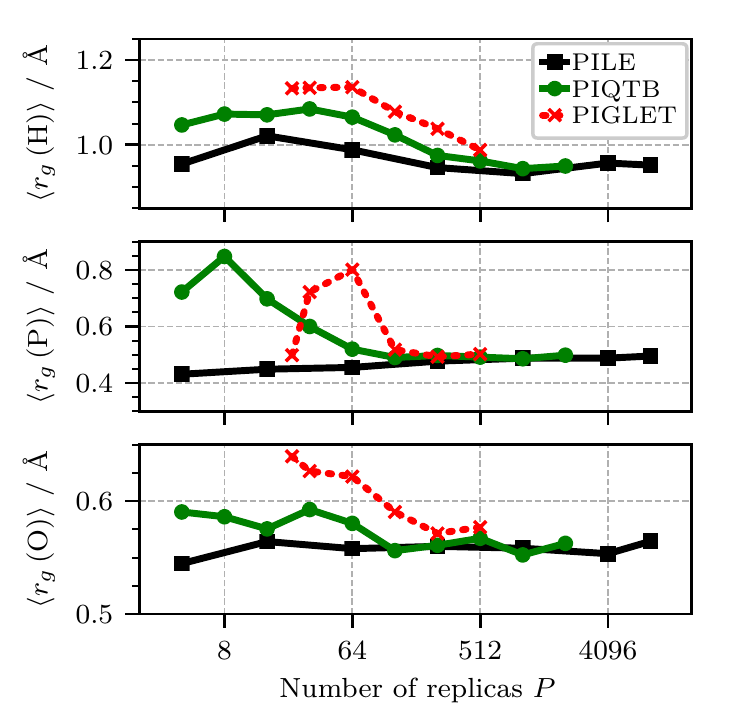}
\caption{Convergence of the radius of gyration $\langle r_{\rm g}\rangle$
of the Zundel cation, \zundel,
         with respect to the number of replicas $P$ for the 
         dangling hydrogen atoms (top), the shared proton (middle)
         and the oxygen atoms (bottom) at \SI{1.67}{\kelvin} as obtained
         using the PILE (black squares), PIGLET (red crosses) and
         PIQTB (green circles) thermostats.
        }
\label{fig:rgyr}
\end{figure}

While the energetic and structural properties both showed only
small changes when decreasing the temperature, quantum delocalization
increases significantly when $T$ decreases.
For example, the radius of gyration of the dangling hydrogens is 
increasing from $r_{\rm g}\approx$~\SI{0.2}{\AA} at \SI{100}{\kelvin} to 
$r_{\rm g}\approx$~\SI{0.9}{\AA} at \SI{1.67}{\kelvin}.
It has recently been shown that the PIGLET method can exhibit a very slow 
Trotter 
convergence of $r_{\rm g}$ at low temperatures and that it can even become
difficult to converge this quantity at ultra-low temperatures~\cite{Uhl2016}.
We thus verify that the PIGLET and PIQTB methods are able
to converge towards the correct value of the radius of gyration.
The dependence of the radius of gyration on the number of path integral
replicas using the three different thermostatting approaches
is depicted in Fig.~\ref{fig:rgyr} for \SI{1.67}{\kelvin}; 
the convergence at the other two temperatures is similar and
can be found in the Supporting Information.
We first note that, compared to the convergence of all the other quantities 
studied here, the convergence of $r_{\rm g}$ is surprisingly fast with 
the PILE thermostat.
Indeed, at \SI{1.67}{\kelvin} convergence is reached roughly 
at $P \approx 256$ replicas for all 
nuclei
in the system.
Both, PIGLET and PIQTB converge towards the correct value of $r_{\rm g}$,
however the convergence is rather slow and the required number of 
beads is approximately the same as for the PILE thermostat.
It is also worth mentioning that the PIQTB and PIGLET methods approach
the converged value from above and, therefore, both methods tend to {\em overestimate}
the delocalization~-- if the number of path integral replicas is {\em too small}.
This overestimation for lower Trotter numbers is caused
by transient exchange of the shared proton by one of the dangling
hydrogen atoms for some of the beads that renders the
unique separation into shared proton and dangling hydrogens
impossible
in this case.
Note that the lowest normal mode frequencies that are directly 
related to the evolution of the radius of gyration evolve rather 
slowly at ultra-low temperatures.
This leads to some minor statistical fluctuations
visible in Fig.~\ref{fig:rgyr} although every simulation was
propagated for \SI{0.5}{\nano\second}. 
Overall, 
these results clearly prove 
that colored noise thermostats do describe
the quantum delocalization of the nuclei correctly
when using a sufficiently large number of replica, 
however without any major acceleration compared to standard PIMD.

\subsection{Anisotropic Shape of the Hydrogen-bonded Proton}
\label{sec:shape}
After having established that colored noise
thermostatting methods are able to correctly describe energetic, structural and
delocalization properties of the Zundel cation 
down to ultra-low temperatures, 
we can use this system as a probe for 
different 
hydrogen-bonding
situations to
draw some more general conclusions. 
For this purpose it is important to note that the
proton transfer barrier in the Zundel cation
is well-known~\cite{Marx2006,*Marx2007} 
to feature a very strong dependence on the oxygen-oxygen distance $r_\text{OO}$.
More interestingly, 
the anisotropy of the nuclear quantum delocalization of such shared protons
has been shown~\cite{Benoit2005} to be strongly dependent on the hydrogen bond length.
\begin{figure}
\includegraphics[width=0.5\textwidth]{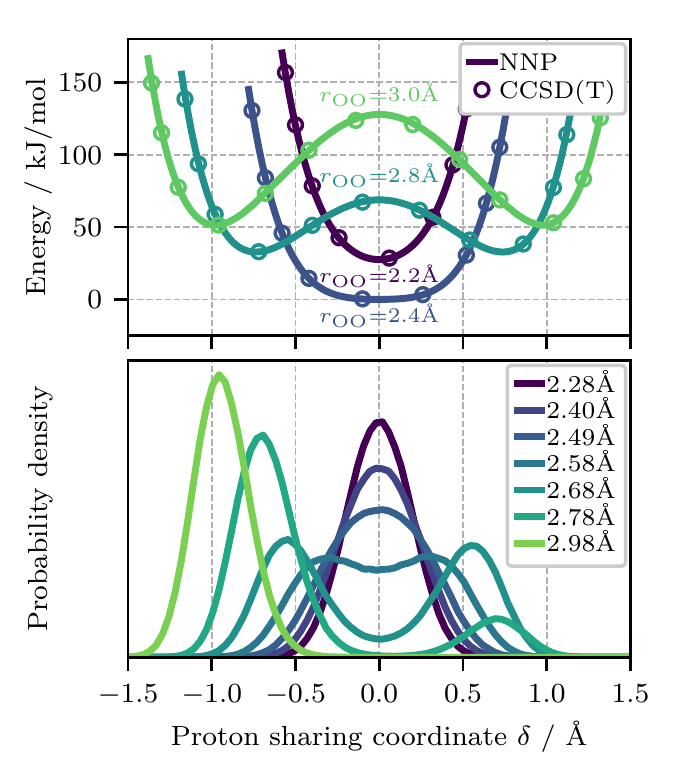}
\caption{Top: Potential energy profile 
of the Zundel cation, \zundel{}, 
along the proton sharing 
(transfer) 
coordinate 
         $\delta$ with the oxygen distance $r_\text{OO}$ constrained at different values
(see text). 
         The line is the energy profile obtained with the Neural Network Potential (NNP)
         and the open circles show the energy as directly obtained from the  
         CCSD(T*)-F12a/AVTZ
         Coupled Cluster electronic structure calculations;
         the energies are 
reported
	 relative to the equilibrium structure of \zundel{}
         and the 
vertical 
         offset is thus the shift of the potential energy surface
         due to constraining the donor-acceptor distance using the reported $r_\text{OO}$ value.
Bottom: Associated distribution functions of the $\delta$ 
         coordinate at \SI{100}{\kelvin} obtained using PILE simulations with $P=1024$ beads
(not symmetrized, see text for simulation protocol and sampling statistics) 
         and $r_\text{OO}$ restrained 
(see Sec.~\ref{sec:comp-det} for details) 
         so that the average 
         oxygen-oxygen distance $\langle r_\text{OO} \rangle$ 
(which is reported in the legend) 
         remains close to the target value. 
} 
\label{fig:delta_doo}
\end{figure}
Fig.~\ref{fig:delta_doo}~(top) 
depicts
cuts through the 
Coupled Cluster PES 
along the proton sharing coordinate $\delta$ 
for different 
donor-acceptor
distances.
For short distances, the potential energy profile 
is characterized by 
a single well centered around the midpoint of the O-O distance.
Increasing $r_\text{OO}$ leads to the appearance and 
growth of a centered barrier that separates two energetically equivalent
proton positions, so that the energy profile 
for proton transfer
becomes a double well potential.  
This typical dependence is encountered in almost every hydrogen bond,
although 
asymmetric environments typically 
lift the perfect degeneracy of the two minima.
We can exploit this to use the Zundel cation
as a prototypical model of different 
hydrogen-bonding
situations by imposing a specific O-O distance,
mimicking strong and weak hydrogen bonds in case of
short and long donor-acceptor distances. 
Indeed, the associated distribution of the proton sharing
coordinate $\delta$ at \SI{100}{\kelvin} for
different restrained values of $r_\text{OO}$
shown in Fig.~\ref{fig:delta_doo} (bottom)
demonstrates that we can describe fully
centered up to very asymmetric 
hydrogen-bonding
situations with this approach.

In the following, we 
assess
in detail 
if colored noise thermostats are able to describe correctly the expected 
anisotropy of the 
nuclear quantum delocalization of the shared
proton~\cite{Benoit2005}
for different values of the oxygen-oxygen distance in \zundel . 
\begin{figure*}[ht]
    \begin{minipage}{0.72\textwidth}
        \includegraphics[width=1.0\textwidth]{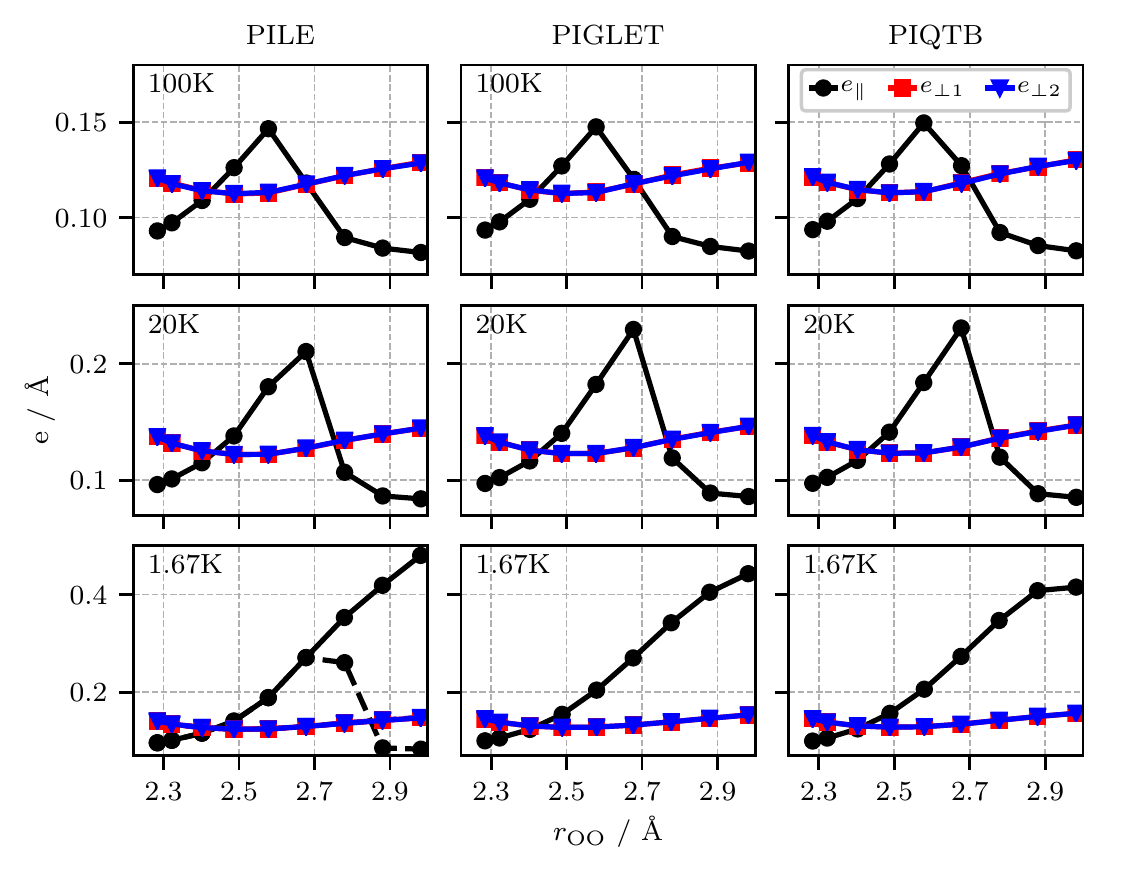}
    \end{minipage}
    \begin{minipage}{0.27\textwidth}
        \includegraphics[width=1.0\textwidth]{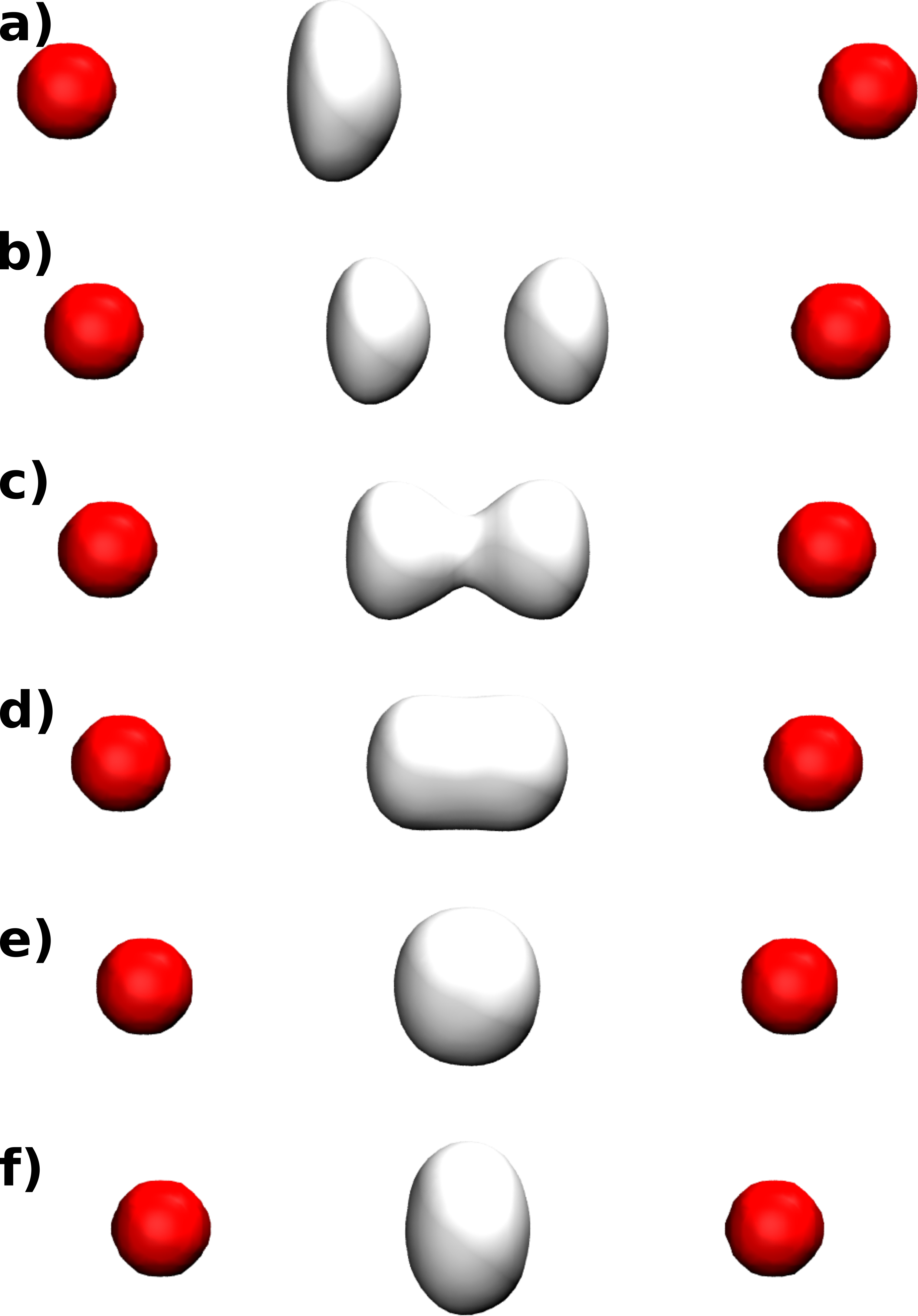}
    \end{minipage}
\caption{
         Left: 
Anisotropic nuclear quantum delocalization of the shared proton
in the Zundel cation, \zundel, 
given by the
extent
         of the ellipsoid along the eigenvectors
         $e_{\parallel},e_{\perp 1},e_{\perp 2}$
         of the gyration tensor
as defined in Eq.~(\ref{eq:gyrationtensor}) 
         as a function of the oxygen-oxygen atom distance from
         restrained simulations using the PILE (left column),
         PIGLET (middle column) and PIQTB (right column) thermostats.
         The eigenvalues have been ordered such that the
         corresponding eigenvector of $e_{\parallel}$ is oriented along
         the O-O vector, while the eigenvectors of $e_{\perp 1}$ and 
         $e_{\perp 2}$ are orthogonal to the O-O vector;
note that the latter two data sets are essentially on top of each other. 
         Simulations at $T=$\SI{100}{}, \SI{20}{} and \SI{1.67}{\kelvin}
         using the PILE thermostat were performed with \num{1024},
         \num{2048} and \num{4096} beads, respectively,
         while PIQTB and PIGLET used \num{32}, \num{128} and \num{512}
         replicas, respectively.
         The dashed line for the PILE results at \SI{1.67}{\kelvin}
         is obtained from non-centered proton centroid simulations 
         as described in the text.
         Right: 
         Quantitative representation 
         of the nuclear densities 
(a.k.a. spatial distribution functions, SDFs)
of the shared
         proton (white) and the oxygen atoms (red) as a function of
         the O-O distance with a) \SI{3.0}{\AA}, b) \SI{2.8}{\AA}, c) \SI{2.7}{\AA},
         d) \SI{2.6}{\AA}, e) \SI{2.4}{\AA}, f) \SI{2.25}{\AA} 
         from PILE simulations at \SI{20}{\kelvin}.
         }
\label{fig:rgyr_vs_doo}
\end{figure*}
This is achieved in practice by 
restraining the 
oxygen atoms by two external harmonic potentials
each acting on one oxygen atom
such that 
the 
quantum expectation
value of $r_\text{OO}$,
i.e. $\langle r_\text{OO} \rangle$,
remains close to the preset target value
while the O-O distance still exhibits some fluctuations;
see Sec.~\ref{sec:comp-det} for implementation details.
This was done for all three temperatures and all thermostatting
methods at nine different O-O distances varying from
2.25 
up to \SI{3.0}{\AA}, thus allowing
us to probe the different proton transfer barriers
shown in Fig.~\ref{fig:delta_doo}
including the barrierless limit. 
In the subsequent analysis we particularly focus on the anisotropy
of the nuclear quantum delocalization of the shared proton in these
various simulations.
This is achieved by calculating the 
quantum expectation value of the 
gyration tensor
\begin{align}
    I_{\rm g}^{\alpha,\beta} = \left< \frac{1}{P} \sum_{s=1}^{P} 
                           \left(\text{r}_s^{\alpha}-\text{r}_{\rm c}^{\alpha}\right)\cdot
                           \left(\text{r}_s^{\beta}-\text{r}_{\rm c}^{\beta}\right)  \right>
, 
\label{eq:gyrationtensor}
\end{align}
where $\text{r}_s^{\alpha/\beta}$ and $\text{r}_{\rm c}^{\alpha/\beta}$ 
are the components of the position of the bead~$s$ 
and of the centroid of the ring polymer associated 
with the shared proton, respectively 
(where $\alpha,\beta = x,y,z$).
Next, by diagonalizing 
the $I_{\rm g}^{\alpha,\beta}$
matrix one can extract quantitative information
on the anisotropy of the nuclear quantum delocalization 
and thus on the general shape of the
proton in the different hydrogen-bonding situations.
The square root of the eigenvalues can be interpreted
as the extent of the nuclear quantum delocalization
in the direction of the corresponding eigenvector.
We chose to order these eigenvalues according to
the orientation of the corresponding eigenvector
with respect to the O-O vector.
The first eigenvalue $e_{\parallel}$ is always 
oriented 
along the O-O vector, while $e_{\perp 1}$ and $e_{\perp 2}$ are orthogonal to it.
As previously established~\cite{Benoit2005}
using a very similar analysis based on the inertia tensor, 
protons in hydrogen bonds 
can exhibit three distinct shapes
depending on the average donor-acceptor distance. 
If $e_{\parallel} > e_{\perp 1} \approx e_{\perp 2}$,
the proton has a ``cigar-like'' shape 
(see density~d on the right side of Fig.~\ref{fig:rgyr_vs_doo}), 
while for $e_{\perp 1} \approx e_{\perp 2} > e_{\parallel}$ 
a disk is formed (distribution~a or~f) 
whereas if $e_{\parallel} \approx e_{\perp 1} \approx e_{\perp 2}$ 
the shape of the proton resembles a sphere (density~e).

The dependence of these eigenvalues on the 
oxygen-oxygen 
distance for the three temperatures and the
different thermostats under study is depicted 
in Fig.~\ref{fig:rgyr_vs_doo}, where the bead numbers
for the different thermostatting methods 
were selected as specified in the caption according to
convergence study from
the previous section.
Let us start by discussing the results obtained
at the highest temperature, \SI{100}{\kelvin}.
The eigenvalues of the gyration tensor
reveal that for 
strong compression, \textit{i.e.} in the limit of extremely strong hydrogen bonds and thus
very short O-O distances (\SI{2.25}{} to \SI{2.3}{\AA}), 
the proton adopts a disk-like shape in the central position between 
the two oxygen atoms indicating that it is ``squeezed''
between the two heavy atoms (see density~f on the right side
of Fig. \ref{fig:rgyr_vs_doo}).
At \SI{2.4}{\AA}, the shape of the proton is
more spherical (corresponding density~e) before becoming cigar-like 
from \SI{2.5}{\AA} to \SI{2.6}{\AA}, thus 
spreading along the O-O vector (see density~d).
For these distances, the proton transfer
barrier is first absent and then
appears but remains small, so that the transfer 
happens essentially as if no barrier was present.
In the latter case, centering of the hydrogen bond 
takes place in a double well potential due to zero-point motion
(as relevant, for instance, on the way of molecular ice~VIII to ionic ice~X
as a result of applying extreme hydrostatic pressures~\cite{Benoit2002}).
Since the potential 
becomes wider upon increasing $r_\text{OO}$, 
the nuclear wavefunction of the proton can stretch
out in the direction of the oxygen atoms which at
some point is assisted by tunneling over the
emerging barrier.
At an O-O separation of \SI{2.7}{\AA}, the barrier is
high
enough that only thermal proton hopping is present 
at \SI{100}{\kelvin}
(\textit{i.e.} only hopping of the whole ring polymer is observed
rather than the polymer being located simultaneously in both wells
which would herald proton tunneling in hydrogen bonds~\cite{Tuckerman2001}). 
This results again in a spherical shape, since
the proton is now mainly located closer to either O$_1$ or O$_2$
but does not spread over the two wells.
For larger distances, however, the shape finally evolves into being disk-like again.
As the barrier height is increasing, the proton
now stays trapped in one of two wells
(\textit{i.e.} no proton transfer events are observed
during the whole simulation given the finite sampling time
of \SI{0.5}{\nano\second}). 
The emergence of the disk-like shape can
be understood by the decreasing attraction
towards the opposite oxygen atom that
leads to a larger extent of the nuclear wavefunction
along the direction perpendicular to the O-O vector.
This dependence of the proton shape on the
oxygen-oxygen
distance is additionally
visualized in Fig.~\ref{fig:rgyr_vs_doo} (right)
where 
also the nuclear densities of the oxygen atoms are depicted to scale with that of the proton. 
One-to-one comparison to the colored noise thermostats demonstrates that
these \SI{100}{\kelvin} PILE results,
all the way from ultra-strong to very weak 
hydrogen-bonding,
are rather well reproduced by 
both, PIGLET and PIQTB approaches
using only a fraction of the number of beads.

Let us next focus on the temperature dependence of these
properties.
At \SI{20}{\kelvin} the general trends described for
\SI{100}{\kelvin} remain unchanged however, the 
O-O region where the proton adapts a cigar-like shape
is now extended up to \SI{2.7}{\AA}.
In addition, the extent of the delocalization
especially in the direction parallel to the O-O vector
increases and has its maximum at \SI{2.7}{\AA}.
This is due to an increase of the tunneling 
contribution to the proton transfer upon lowering 
the temperature that counteracts the decrease 
of thermal fluctuations.
In practice, here we go from a mainly thermally activated
proton transfer scenario at $T=$~\SI{100}{\kelvin} 
to a regime where tunneling is the dominating 
process at $T=$~\SI{20}{\kelvin} 
(\textit{i.e.} the ring polymer is spreading 
over the two wells
rather than hopping as a compact ball from one well to the
other as observed upon thermal activation)
for this specific O-O separation.
Again, both PIGLET and PIQTB are able to
reproduce the PILE results, however slightly
overestimating the delocalization 
of the wavefunction 
along the hydrogen bond, $e_{\parallel}$,  
for \SI{2.7}{\AA}.

Upon lowering the temperature further 
one expects that the tunneling contribution
increases even more.
Due to the degeneracy of the two minima 
of the double well potential,
the ground state wavefunction
is required to be symmetric
with respect to the center of the O-O distance
and we should thus always obtain some tunneling contributions
(even if very small when the barrier width and height are large).
At ultra-low temperatures one should therefore 
expect that the region with a cigar-like proton
shape, 
being
associated with tunneling, 
further expands towards larger O-O distances.
This is exactly what is observed for PIGLET and PIQTB
at \SI{1.67}{\kelvin} where we obtain a cigar-like shape of the proton 
even at 
donor-acceptor distances as large as
\SI{3.0}{\AA}.
In practice, however, the proton can still be trapped
on one side of the molecule, 
due to non-ergodic sampling of the path integral, 
if the barrier is too high.
This effect can be seen for the PILE results at \SI{1.67}{\kelvin}
in Fig.~\ref{fig:rgyr_vs_doo} (dashed lines),
where the proton has been initialized closer to one 
of the oxygen atoms
according to our standard protocol. 
Even within
our simulation length of \SI{0.5}{\nano\second} we
are not able to build up significant tunneling contributions
for \SI{2.9}{} and \SI{3.0}{\AA} and only partial tunneling
is observed 
at \SI{2.8}{\AA} 
when using the PILE thermostat.
As already mentioned for the radius of gyration, at 
such ultra-low
temperatures,
the lowest normal mode frequencies of the ring polymer
are very important for the correct description of 
delocalization and tunneling.
Since these modes evolve rather slowly, we see 
non-ergodic
trapping of the proton at one side
of the molecule for the PILE simulations at \SI{1.67}{\kelvin}
and O-O distances larger than \SI{2.8}{\AA}.
However, if we initialize the proton in our simulations
in the central position between the oxygen atoms and
additionally restrain the centroid 
position to stay around the center of the O-O 
vector by a very soft harmonic potential
being \num{25000} times weaker than the external
harmonic potentials acting on the oxygen atoms
(providing the solid line for PILE at \SI{1.67}{\kelvin}
in Fig.~\ref{fig:rgyr_vs_doo}), 
we can enforce a symmetric distribution.
By doing so, we are able to obtain the same dependence of
the eigenvalues of the gyration tensor 
as a function of
the O-O distance for PILE 
as observed for PIGLET and PIQTB.
This analysis reveals that the 
colored noise thermostatting schemes
tend to overestimate the tunneling contribution
at these temperatures.
However, due to inefficient sampling, standard PIMD simulations
such as PILE 
have a tendency to underestimate these contributions 
especially at ultra-low temperatures and large O-O distances.

In summary, we showed that the 
anisotropic
shape of the proton
in the Zundel cation depends 
exquisitely
on the 
donor-acceptor distance and thus on
hydrogen bond length
and, moreover, features exactly the same dependence as 
discovered 
in Ref.~\citenum{Benoit2005} for ice at different 
pressures.
As expected, the tunneling regime that is
corresponding to a cigar-like shape of the
proton extends towards larger O-O separations,
if the temperature is lowered.
In addition, this anisotropy of the shape of the
proton in
hydrogen-boded
system is correctly
reproduced by both PIGLET and PIQTB
even down to ultra-low temperatures.

\section{Conclusions and Outlook}
\label{sec:conclusion}

In conclusion, we have shown that 
two modern
colored noise thermostatting methods
that have been devised to accelerate path integral
sampling using molecular dynamics techniques, 
namely PIGLET and PIQTB, 
are able to speed up the convergence of
energetic and structural properties of the Zundel 
cation, \zundel, 
even down to ultra-low temperatures on the order of one~Kelvin,
although highly anharmonic properties such as the proton sharing coordinate
need to be treated with care.
Upon suitably restraining the oxygen-oxygen distance,
and thus the length of the hydrogen bond, 
from very short to fairly long donor-acceptor distances, 
we are able to drive this prototypical system
from ultra-strong to very weak
hydrogen-bonding
scenarios.
The latter are known to lead to greatly different
nuclear quantum effects and thus delocalization as well as tunneling
properties of the shared proton depending on temperature. 
Using Behler's Neural Network Potentials of
essentially converged Coupled Cluster accuracy, 
CCSD(T*)-F12a/AVTZ,
enables us to perform path integral simulations using
the ``gold standard'' of quantum chemistry for describing 
the interactions of this hydrogen-bonded complex.

Both, PIGLET and PIQTB are shown to 
converge to the correct nuclear 
delocalization properties like the radius of gyration.
Moreover, 
both methods are able to correctly
reproduce the anisotropy of the shape
of the shared proton in all possible 
hydrogen-bonding
situations,
although they slightly but systematically overestimate tunneling contributions. 
Overall, our results validate that
the use of these colored noise thermostatting methods
can provide great advantage over standard PIMD
simulations
by reducing the Trotter number that is required for the desired
level of quantum convergence.

Given the generic character of the studied hydrogen bond depending
on its length and on temperature, we expect
that our conclusions are valid much beyond the \zundel\ complex
and should therefore apply to 
nuclear quantum effects on 
hydrogen bonding in general.
This includes, in particular, 
hydrogen-bonded
systems at ultra-low temperatures
as for example encountered in superfluid helium
environments
or in He-tagged molecular complexes.

\begin{acknowledgments}
It gives us great pleasure to thank Harald Forbert and 
Felix Uhl for helpful discussions.
This research is part of the Cluster of Excellence “RESOLV”
(EXC 1069) funded by the \textit{Deutsche Forschungsgemeinschaft}, DFG.
C.S. acknowledges partial financial support from the 
\textit{Studienstiftung des Deutschen Volkes} as well as from the
\textit{Verband der Chemischen Industrie}.
The computational resources were provided by HPC@ZEMOS,
HPC-RESOLV, BOVILAB@RUB, and RV-NRW.
\end{acknowledgments}
\section*{Supporting Information}
See the supplementary material for additional details on the
implementation of the PIQTB into \texttt{CP2k}
and supporting analyses of structural and nuclear delocalization
properties.
\end{document}